\documentclass[prl,twocolumn,amssymb,amsmath]{revtex4-1}
\usepackage{graphicx}
\usepackage{epsfig,color}
\usepackage{amsmath,amssymb}
\begin{document}

\title{Confinement and lattice QED electric flux-tubes simulated with ultracold atoms}

\date{\today}

\author{Erez Zohar}
\affiliation{School of Physics and Astronomy, Raymond and Beverly Sackler
Faculty of Exact Sciences, Tel-Aviv University, Tel-Aviv 69978, Israel.}
\author{Benni Reznik}
\affiliation{School of Physics and Astronomy, Raymond and Beverly Sackler
Faculty of Exact Sciences, Tel-Aviv University, Tel-Aviv 69978, Israel.}

\begin{abstract}
We propose a method for simulating 2+1-d compact lattice quantum-electrodynamics (QED), using ultracold atoms in optical lattices.
In our model local Bose-Einstein condensates' phases correspond to the electromagnetic vector-potential, and the local number operators
represent the conjugate electric field.
The well-known gauge invariant Kogut-Susskind Hamiltonian is obtained as an effective low energy theory. The field is then coupled to
external static charges.
We show that in the strong coupling limit this gives rise to 'electric flux-tubes'
and to confinement. This can be observed by measuring the local density deviations of the BECs, and is expected to hold even, to some extent,
outside the perturbative calculable regime.
\end{abstract}

\maketitle

Free quarks are not found in nature; This is due to the mechanism
of \emph{confinement}. A lot of theoretical progress in this area
has been achieved - either in the lattice Euclidean approach
by Wilson \cite{Wilson}, in nonperturbative methods by Polyakov \cite{Polyakov} or using the
lattice Hamiltonian formalism, by Kogut and Susskind \cite{KogutSusskind, KogutLattice}.

Although gauge theories can be latticized either in a compact (nonlinear) or noncompact (linear) manner,
the compactness is essential to the confinement mechanism \cite{DrellQuinnSvetitskyWeinstein}.
It has been shown that in an abelian 3+1 compact lattice gauge theory, a phase transition
is supposed to take place between two phases - the Coulomb phase
for small couplings, which exhibits the 'regular' $V\left(R\right)\propto1/R$
static potential between two $R$-separated static charges, and the
confining phase, for which the static potential is linear in the distance
between the charges - $V\left(R\right)\propto R$, for large values
of the coupling constant \cite{KogutLattice}. (Nonabelian theories, on the other hand, confine for all
values of the coupling constant). However, for an abelian 2+1 compact lattice gauge theory, confinement
was shown to take place for all the values of the coupling constant, due to nonperturbative effects of instantons
\cite{Polyakov,BanksMyersonKogut,DrellQuinnSvetitskyWeinstein,BenMenahem}.
Considering thermal effects as well, even in 2+1 dimensions a phase transition to a Coulomb phase exists
for $T>0$ \cite{Svetitsky,SvetitskyReview}.

The mechanism responsible for confinement is believed to produce an 'electric flux tube', connecting two static charges
in the confining phase, which is hard to measure directly. It requires measuring the force
and/or potential between two static charges. If one wishes to observe the phase transition, the coupling constant
has to be varied, which poses another difficulty. A quantum simulation of such a model could
allow a direct test of the confinement mechanism and the phase transitions.

Quantum gases of ultracold atoms, implemented in optical lattices \cite{Bloch}, provide models with highly controllable
parameters and offer a natural playground for the simulations of such models.
Quantum simulation approaches of various kinds and aspects of compact U(1) pure gauge theory, in cold gases and other systems,
have been proposed by several authors:
In \cite{Pyrochlore}, an effective theory of U(1) spin liquid in pyrochlore was discussed; in \cite{RingExchange},
using a molecular state in optical lattices, an effective theory of ring exchange was derived, and it is, in the limit of no hopping,
a U(1) lattice gauge theory, with a Coulomb phase; in \cite{ArtificialPhotons}, emergence of 'artificial photons'
and a Coulomb phase in an effective theory based on dipolar bosons in an optical lattice were shown; and in \cite{Rydberg},
a possibility to simulate a spin U(1) pure gauge theory as a low energy theory with a system of Rydberg atoms was presented.

In this letter, we suggest a method for simulating compact QED with cold atoms in optical lattices, which should enable
a direct observation of 'electric flux tubes' that emerge in the mechanism of confinement.
In our model, the vector potential and its conjugate electric field
are represented by the local condensate phase operators and their conjugate number operators.
These observables 'live' on the links of a two- or three-dimensional optical lattice, and hence each link of the lattice
is here represented by a separate Bose-Einstein condensate.
In order to obtain the QED Hamiltonian, one has to generate certain two-  and four-body
interactions between the condensates, that manifest local gauge invariance.
In order to avoid the hopping processes of an ordinary Bose-Hubbard model, we introduce a four-species two dimensional set-up (fig. 1).
The condensates are located on the links of a lattice -  each species on a different link - and overlap at the lattice's vertices.
Hence, condensates of the same type are spatially separated, as depicted in figure 1, causing the attenuation of hopping processes.
Next we use Raman transitions and two atom scattering processes in order to create special 'diagonal' hopping and nonlinear interactions.
As we show, in this new setup a certain choice of parameters gives rise to gauge invariance in the low energy sector, hence
compact QED emerges as an effective theory.

To study the effect of confinement within this setup, we can introduce two spatially separated effective 'charges' by creating local deformations of
the trapping potential at the position of the 'charges' at the relevant vertices. We then expect that the local atomic densities, within the QED parameter
regime should manifest the effect of confinement by the appearance of a flux-like tube of alternating atomic density deviations along the
line connecting the 'charges' (fig. 2), while such a 'flux tube' will not appear outside the QED parameters regime.
Other possible implications of our model will be shortly discussed in the summary.

\begin{figure}
  \includegraphics[scale=0.35]{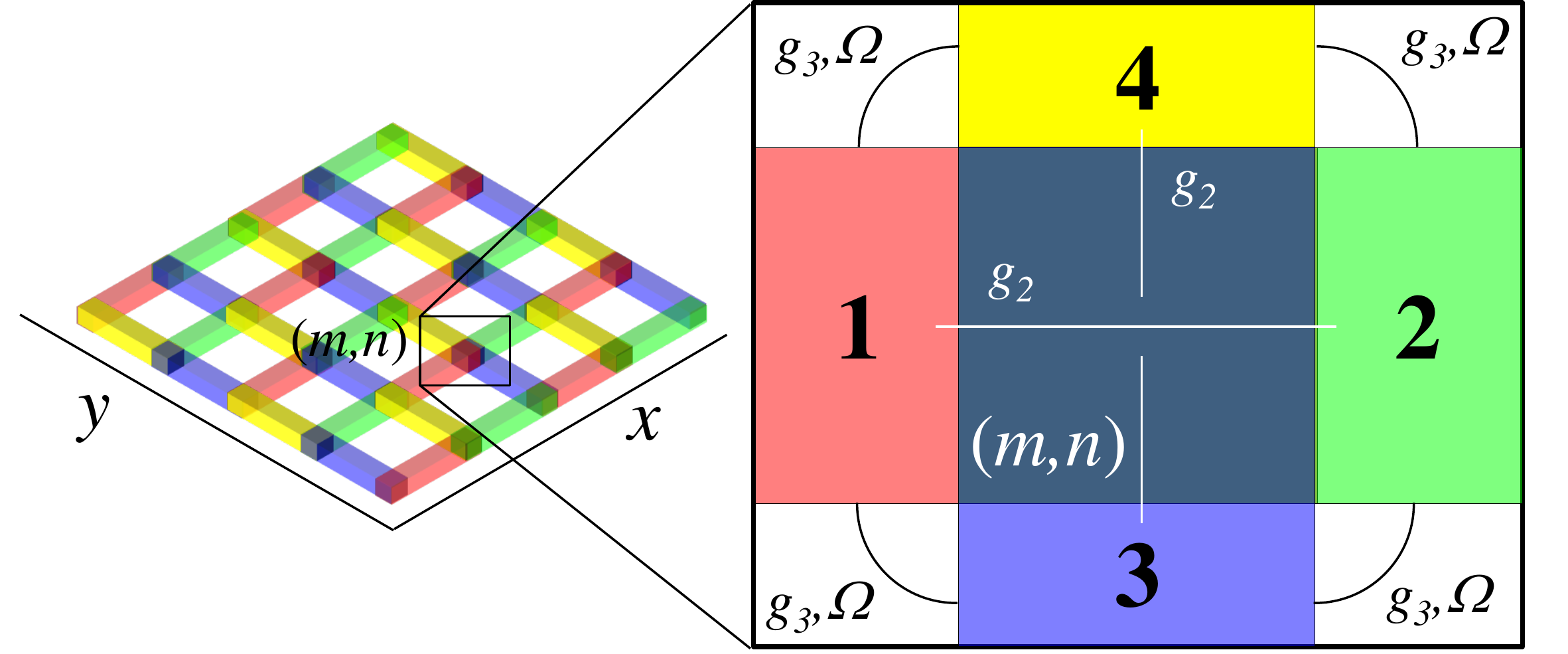}\\
  \caption{On the left - structure of the lattice. The different condensate species are colored in four colors;
  the colored boxes represent the links (condensates), and there the localized wavefunctions are concentrated.
  At the vertex (symbolized by a cube) the wavefunctions of the neighboring links overlap and these are the only overlap integrals which are not negligible.
  On the right - a close-up picture of a single vertex, showing the various interaction parts of the Hamiltonian -
  scattering and hopping.}\label{fig1}
\end{figure}

We bgein with a system of condensates described by the Hamiltonian
$H=\int d^{3}x \underset{i,j=1}{\overset{4}{\sum}}\mathcal{H}_{ij}\left(\mathbf{x}\right)$,
where
\begin{widetext}
\begin{equation}
\mathcal{H}_{ij}\left(\mathbf{x}\right)=\Psi_{i}^{\dagger}\left(\mathbf{x}\right)\Big(\delta_{ij}
\left(\mathcal{H}_{0}^{i}\left(\mathbf{x}\right)+V_{M}(\mathbf{x})\right)+\Omega_{ij}\Big)\Psi_{j}
\left(\mathbf{x}\right)+\frac{g_{ij}}{2}\Psi_{i}^{\dagger}\left(\mathbf{x}\right)\Psi_{j}^{\dagger}
\left(\mathbf{x}\right)\Psi_{j}\left(\mathbf{x}\right)\Psi_{i}\left(\mathbf{x}\right) \label{eq:H0}
\end{equation}
\end{widetext}
$\delta_{ij}$ is Kronecker's Delta, $g_{ij}$ are the s-wave scattering coefficients and $\Omega_{ij}$ are Rabi frequencies. It contains the following parts:
\emph{The 'free' Hamiltonian} of each 
species:
 $\mathcal{H}_{0}^{i}\left(\mathbf{x}\right)=-\frac{\nabla^{2}}{2m}+V_{i}\left(\mathbf{x}\right)$, where $V_{i}\left(\mathbf{x}\right)$ is the optical lattice trapping potential of the species $i$;
\emph{the scattering terms}, set by the coupling constants $g_{ij}$ (neglecting the three- and four-body interactions):
(i) self-scattering terms - $g_{ii}\equiv g_{1}$,
(ii) two-species scattering terms: $g_{12}=g_{21}=g_{34}=g_{43} \equiv g_{2}$ along straight lines,
and along the diagonals, $g_{13}=g_{31}=g_{14}=g_{41}=g_{23}=g_{32}=g_{24}=g_{42} \equiv g_3$ (all the other $g_{ij}$'s are zero);
an \emph{'external charges' simulating potential}, which deforms the lattice potential at the vertices and is approximated by a very localized potential:
$V_{M}(\mathbf{x}) \equiv \underset{m,n}{\sum}\alpha_{m,n}\delta\left(\mathbf{x}-\mathbf{x}_{m,n}\right)$, where $\alpha_{m,n}$
are constants (whose value and sign are related to the 'external charges' and will be determined in the sequel) and $\mathbf{x}_{m,n}$ is the position of the $(m,n)$ vertex.
The \emph{laser generated 'Rabi terms'}
$\Omega_{13}=\Omega_{31}=\Omega_{14}=\Omega_{41}=\Omega_{23}=\Omega_{32}=\Omega_{24}=\Omega_{42}\equiv\Omega'_{0}$
couple the condensates to each other in a special, 'diagonal' manner, as depicted in figure \ref{fig1}. (All the other $\Omega_{ij}$'s are zero).
Since the minima of the same species are far enough apart, the hopping effects are solely controlled by the latter 'Rabi terms'.
Experimentally,
our scheme can be implemented by using
holographic masks  techniques  \cite{OpticalManipulationReview} in order to generate the required optical lattice and using optical
Feshbach resonances in order to control
the coupling strengths $g_{ij}$ \cite{Fedichev,Bohn,Fatemi}).
Raman transitions can be used to control the coefficients $\Omega_{ij}$ of the 'Rabi terms'.

The second quantization wavefunctions of the condensates (taking into account only the lowest band excitations) are
$\Psi_{1,2}=\underset{m,n}{\sum}a_{m,n}\phi_{m,n}\left(\mathbf{x}\right),
\Psi_{3,4}=\underset{m,n}{\sum}b_{m,n}\chi_{m,n}\left(\mathbf{x}\right)$
where $a_{m,n},b_{m,n}$ are single-mode annihilation
operators, annihilating one particle in the ground state of the corresponding
link (minimum). Note that because of the lattice's structure,
not all the values $(m,n)$ are included in the wavefunction of each species.
We assume that the local Wannier functions \cite{Bloch} respect the symmetries
$\phi_{m,n}\left(\mathbf{x}\right)=\phi\left(\mathbf{x-x}_{m,n}^{\mathrm{1,2}}\right),\chi_{m,n}^{\mathit{}}\left(\mathbf{x}\right)=
\chi\left(\mathbf{x-x}_{m,n}^{\mathrm{3,4}}\right)=\phi\left(R\mathbf{x}\right)$,
where $R$ is the appropriate rotation operator, and that they are chosen to be
real \cite{Wannier}.

Plugging the wavefunctions into the Hamiltonian (\ref{eq:H0}),
one gets, using the above assumptions, that the only non-negligible contributions are:
$2\lambda+\mu\equiv\frac{g_{1}}{2}\int d^{3}x\left|\phi\left(\mathbf{x-x}_{0}\right)\right|^{4}$,
$V_{0}\equiv\int d^{3}x\phi^{*}\left(\mathbf{x-x}_{0}\right)\left(-\frac{\nabla^{2}}{2m}+V_{i}\left(\mathbf{x}\right)\right)
\phi\left(\mathbf{x-x}_{0}\right)$,
$V_{2}\equiv\int d^{3}x\left|\phi\left(\mathbf{x-x}_{0}\right)\right|^{2}\left|\phi\left(\mathbf{x-x}_{1}\right)\right|^{2}$,
$V_{3}\equiv\int d^{3}x\left|\phi\left(\mathbf{x-x}_{0}\right)\right|^{2}\left|\chi\left(\mathbf{x-x}_{2}\right)\right|^{2}$,
$\Delta_{m,n}\equiv-\frac{1}{2\lambda}\alpha_{m,n}\left|\phi\left(\mathbf{x}_{m,n}\right)\right|^{2}$ and
$\Omega_{0}\equiv\Omega'_{0}\int d^{3}x \phi^{*}\left(\mathbf{x-x}_{0}\right)\chi\left(\mathbf{x-x}_{2}\right)$
(Here the reality of the Wannier functions is employed).
$\mathbf{x}_0$ is the position of an arbitrary minimum of the potential (due to the symmetries),
$\mathbf{x}_1$ is an adjacent minimum in the same direction (separated by a single lattice spacing)
and $\mathbf{x}_2$ is an adjacent minimum in the orthogonal direction (rotated).
In the following we assume that $g_{2},g_{3}$ satisfy the relation
$g_{2}V_{2}=g_{3}V_{3}=2\lambda$.

Let $ N_{m,n}^{k} $ be the local number operators,
emanating from the vertex $(m,n)$: for horizontal $(\widehat{\mathbf{x}})$ links $k=1$
and for vertical $(\widehat{\mathbf{y}})$ links $k=2$.
$N_{T}=\underset{m,n,k}{\overset{}{\sum}}N_{m,n}^{k}$, the total number of
particles, is a constant of motion. We choose a subspace by fixing
$N_{T}=\mathcal{N}_{L}N_{0}$, where $\mathcal{N}_{L}$ is the number of links and $N_{0}\gg1$).
Defining $M_{m,n}=4N_{0}+\Delta_{m,n}$, $G_{m,n}=N_{m,n}^{1}+N_{m,n}^{2}+N_{m-1,n}^{1}+N_{m,n-1}^{2}-M_{m,n}$,
after some algebra, one obtains the Hamiltonian
$H_{0}\equiv H-H_{R}=\lambda\underset{m,n}{\sum}G_{m,n}^{2}+\mu\underset{m,n,k}{\sum}(N_{m,n}^{k})^{2}$.
The nearest-neighbor hopping part, which results here from the 'Rabi terms', can be written as
$H_{R}\equiv\Omega_{0}\underset{m,n}{\sum}(a_{m,n}b_{m,n}^{\dagger}+a_{m,n}b_{m+1,n}^{\dagger}+a_{m,n+1}b_{m,n}^{\dagger}+a_{m,n+1}b_{m+1,n}^{\dagger}+h.c.)$.

\emph{Gauss's Law.} We wish to obtain a gauge-invariant theory,
and hence we would like to constrain Gauss's law on the system. This is satisfied in the 'QED regime':
$\lambda\gg\mu$ and $\lambda\gg\Omega_{0}$, in which $H_{R}$ can
be treated as a small perturbation. Let us first find the ground state
of $H_{0}$. After expanding the number operator on each link around $N_{0}$: $N_{m,n}^{k}=N_{0}+\delta_{m,n}^{k}$,
one obtains at each vertex:
$G_{m,n}=\delta_{m,n}^{1}+\delta_{m,n}^{2}+\delta_{m-1,n}^{1}+\delta_{m,n-1}^{2}-\Delta_{m,n}$.
Within the subspace of a constant, conserved $N_{T}$, $\underset{m,n,k}{\sum}\delta_{m,n}^{k}=0$.
Neglecting constants of motion, one can rewrite the Hamiltonian in terms of $\delta_{m,n}^{k}$:
$H_{0}=\lambda\underset{m,n}{\sum}G_{m,n}^{2}+\mu\underset{m,n,k}{\sum}(\delta_{m,n}^{k})^{2}\equiv H_{G}+H_{E}$.
$\lambda\gg\mu$, and hence one would like to minimize $H_{G}$ first.
Thus we get that in the ground state, the sum of $\delta_{m,n}^{k}$'s
around each vertex equals the $\Delta_{m,n}$ of the vertex:
This imposes a 'modified Gauss's Law' (sum instead of discrete
divergence), and hence the $\Delta_{m,n}$'s must
be integers (positive, zero or negative) - this can be set by adjusting the values of the $\alpha_{m,n}$'s in $V_{M}(\mathbf{x})$.
Next, to minimize the entire $H_{0}$ (including $H_{E}$) we would like to choose the
lowest $\delta_{m,n}^{k}$'s which satisfy this constraint.

Define the sublattices $A=\{(m,n):m+n=\textrm{even}\}$, $B=\{(m,n):m+n=\textrm{odd}\}$.
Note that for states that respect Gauss's Law (for which $G_{m,n}\left|\psi\right\rangle =0$),
which will later be the physically interesting states,
the sum of $\Delta_{m,n}$'s of each sublattice must be zero. This
follows from adding the $G_{m,n}$'s of each sublattice, taking into
account that the total particle number deviation is zero.

\emph{Quantum Rotor Approximation.} If we set that at each vertex $\left|\Delta_{m,n}\right|\ll N_{0}$,
we get that on each link, in the ground state of $H_{0},$ within
our subspace, $\left|\left\langle \delta_{m,n}^{k}\right\rangle \right|\ll N_{0}$,
and thus, after taking into account the perturbative corrections,
one obtains that on each link $\sigma_{m,n}^{k}\equiv\left\langle \left(\delta_{m,n}^{k}\right)^{2}\right\rangle^{1/2} \ll N_{0}$.
Note that $\sigma_{m,n}^{k}\ll\frac{\lambda N_{0}\left(N_{0}-1\right)}{\Omega_{0}}$
(because $\frac{\Omega_{0}}{\lambda}\ll N_{0}$), and hence the two
conditions of \cite{Rotor} are fulfilled and the Hamiltonian can
be approximated as a quantum rotor Hamiltonian. $H_{G},H_{E}$ remain
the same, because they are already written in the number deviations'
notation. Because of the phase-number relation
of the condensates, $\left[N_{m,n}^{i},\theta_{m',n'}^{j},\right]=i\delta_{mm'}\delta_{nn'}\delta_{ij}$,
$\left[N_{m,n}^{i},e^{\pm i\theta_{m,n}^{i}}\right]$=$\mp e^{\pm i\theta_{m,n}^{i}}$ and
therefore we can define phase-only lowering and raising operators,
$\widetilde{a}_{m,n}=e^{i\theta_{m,n}^{1}},\widetilde{a}_{m,n}^{\dagger}=e^{-i\theta_{m,n}^{1}},\widetilde{b}_{m,n}=e^{i\theta_{m,n}^{2}},\widetilde{b}_{m,n}^{\dagger}=e^{-i\theta_{m,n}^{2}}$,
which operate on the local number deviations: $\widetilde{a}_{m,n}\left|\delta_{m,n}^{1}\right\rangle =\left|\delta_{m,n}^{1}-1\right\rangle$
etc., and since $N_{0}\gg1$, $\sqrt{N_{0}\left(N_{0}+1\right)}\approx N_{0}$. Thus one gets
$H_{R}=\Omega\underset{m,n}{\sum}(\widetilde{a}_{m,n}\widetilde{b}_{m,n}^{\dagger}+\widetilde{a}_{m,n}\widetilde{b}_{m+1,n}^{\dagger}+\widetilde{a}_{m,n+1}\widetilde{b}_{m,n}^{\dagger}+\widetilde{a}_{m,n+1}\widetilde{b}_{m+1,n}^{\dagger}+h.c.)$,
where $\Omega=\Omega_{0}N_{0}$.

\vspace*{0.1cm}

\emph{Effective Hamiltonian.} Let us look again at the eigenstates and eigenvalues of $H_{G}$.
Since $\left[H_{G},H_{E}\right]=0$, the two Hamiltonians can be mutually
diagonlized. The eigenstates of $H_{E}$ are number states, and we
shall use this basis to diagonalize $H_{G}$ as well. Since $\lambda\gg\mu,\Omega$,
the Gauss Hamiltonian $H_{G}$ is much stronger than the other two,
and therefore one can obtain an effective low-energy theory perturbatively \cite{Cohen_Tannoudji_atomphoton}.
It is physically reasonable to derive an effective Hamiltonian by projecting
to the ground state manifold of $H_{G}$. Let us denote this manifold
by $M$: $M=\left\{ \left|M_{\alpha}\right\rangle :H_{G}\left|M_{\alpha}\right\rangle =0\right\}$.
One can see that it is the physical subspace of states which respect Gauss's law.
The perturbative expansion to second order leads to $H_{eff}=H_{E}+H_{B}$,
where $H_{B}=-\frac{2\Omega^{2}}{\lambda}\underset{m,n}{\sum}
\left(\widetilde{a}_{m,n+1}^{\dagger}\widetilde{b}_{m,n}\widetilde{a}_{m,n}^{\dagger}\widetilde{b}_{m+1,n}+h.c.\right)$ is the desired
gauge invariant four-body plaquette interaction.

\emph{Compact QED analogy.} We next relate this model to compact QED and discuss
the implications. First, let us switch to QED-like variables. In order to do so,
we perform the transformation: $E_{m,n}^{k}\equiv(-1)^{m+n}\delta_{m,n}^{k}$,
$Q_{m,n}\equiv(-1)^{m+n}\Delta_{m,n}$ and $\theta_{m,n}^{k} \rightarrow (-1)^{m+n}\theta_{m,n}^{k}$.
Because of the transformation of the phases of links emanating from sublattice $B$ vertices, these links' raising and lowering
operators have to be swapped.

This transforms $H_{E}$, which can be identified as the 'electric Hamiltonian', to $H_{E}=\mu\underset{m,n,k}{\sum}\left(E_{m,n}^{k}\right)^{2}$,
and
$H_{B}=-\frac{2\Omega^{2}}{\lambda}\underset{m,n}{\sum}\left(\widetilde{a}_{m,n}^{\dagger}\widetilde{a}_{m,n+1}\widetilde{b}_{m+1,n}^{\dagger}\widetilde{b}_{m,n}+h.c.\right)=
-\frac{4\Omega^{2}}{\lambda}\underset{m,n}{\sum}\cos\left(\theta_{m,n}^{1}+\theta_{m+1,n}^{2}-\theta_{m,n+1}^{1}-\theta_{m,n}^{2}\right)$
is the magnetic part of the compact QED Hamiltonian (the cosine's argument is the discrete curl of $\theta_{m,n}$, which is the magnetic field).
Thus we obtained an effective low-energy theory whose Hamiltonian is the compact QED Hamiltonian, constrained with Gauss's law
(which is the low-energy constraint):
\begin{equation}
G_{m,n}|\psi\rangle=(-1)^{m+n}(\textrm{div}\textbf{E}_{m,n}-Q_{m,n})|\psi\rangle=0
\end{equation}

\emph{Confinement of external static charges.} Define a new \emph{finite} energy scale,
$U_{0}=\frac{2}{g^{2}}\mu=\frac{4\Omega^{2}g^{2}}{\lambda}$,
and rescale the Hamiltonian to
\begin{widetext}
\begin{equation}
\overline{H}\equiv H_{eff}/U_{0}=\frac{g^{2}}{2}\underset{m,n,k}{\sum}\left(E_{m,n}^{k}\right)^{2}-\frac{1}{g^{2}}\underset{m,n}{\sum}\cos\left(\theta_{m,n}^{1}+\theta_{m+1,n}^{2}-\theta_{m,n+1}^{1}-\theta_{m,n}^{2}\right)
\end{equation}
\end{widetext}
which is the well-known Kogut-Susskind Hamiltonian for an abelian lattice gauge
theory \cite{KogutSusskind,KogutLattice}. From the definition
of $U$, one gets $g^{4}=\frac{\lambda\mu}{2\Omega^{2}}$.
This Hamiltonian has two limits:
(i) The strong coupling limit: $g\gg1$, or $\frac{\mu}{\Omega}\gg\frac{\Omega}{\lambda}$.
In this limit, we can treat $H_{B}$ as a perturbation to $H_{E}$, and
(ii) The weak coupling limit: $g\ll1$, or $\frac{\mu}{\Omega}\ll\frac{\Omega}{\lambda}$.
In this limit, we can treat $H_{E}$ as a perturbation to $H_{B}$.
In a 3+1 theory, the strong coupling limit is within the confining
phase and the weak coupling limit is within the Coulomb phase, and
a phase transition is expected in between \cite{Wilson,KogutSusskind,Polyakov,KogutLattice,DrellQuinnSvetitskyWeinstein}.
In a 2+1 theory, there is no phase transition and confinement is expected
to occur for all $g>0$ \cite{DrellQuinnSvetitskyWeinstein,BenMenahem}.

The 'external charges' are limited
by the restrictions imposed by the constraint $\underset{\left(m,n\right)\in A}{\sum}\Delta_{m,n}=0,\underset{\left(m,n\right)\in B}{\sum}\Delta_{m,n}=0$.
subtracting the second constraint from the first, one gets $\underset{m,n}{\sum}Q_{m,n}=Q_{total}=0$.
Thus, the total charge has to be zero.
 If we add these constraints, we get another constraint , $\underset{m,n}{\sum}\Delta_{m,n}=0$.  This constraint does not seem to have a QED analogy, but it has to be satisfied in our model.

Consider the case of a system with two unit 'external charges', in the
strong coupling limit. Thus we seek for the ground state configurations
of $H_{E}$ and treat $H_{B}$ as a perturbation. These 'charges' must be of opposite signs, in order to satisfy the charge
restrictions. For simplicity, we assume that the 'charges' are
fixed at the vertices $\left(m,n\right)$ and $\left(m+R,n\right)$. If $R=1$, the 'charges'
are fixed at two vertices of different sublattices, and hence have to have the same sign in terms of $\Delta_{m,n}$. This, however, does not satisfy the charge restrictions, and one has to add more 'charges' to the system. This is true, in fact, for any odd $R$. Therefore we shall consider only the case of an even $R$,
for two 'charges' in the system.

\begin{figure}
  \includegraphics[scale=0.45]{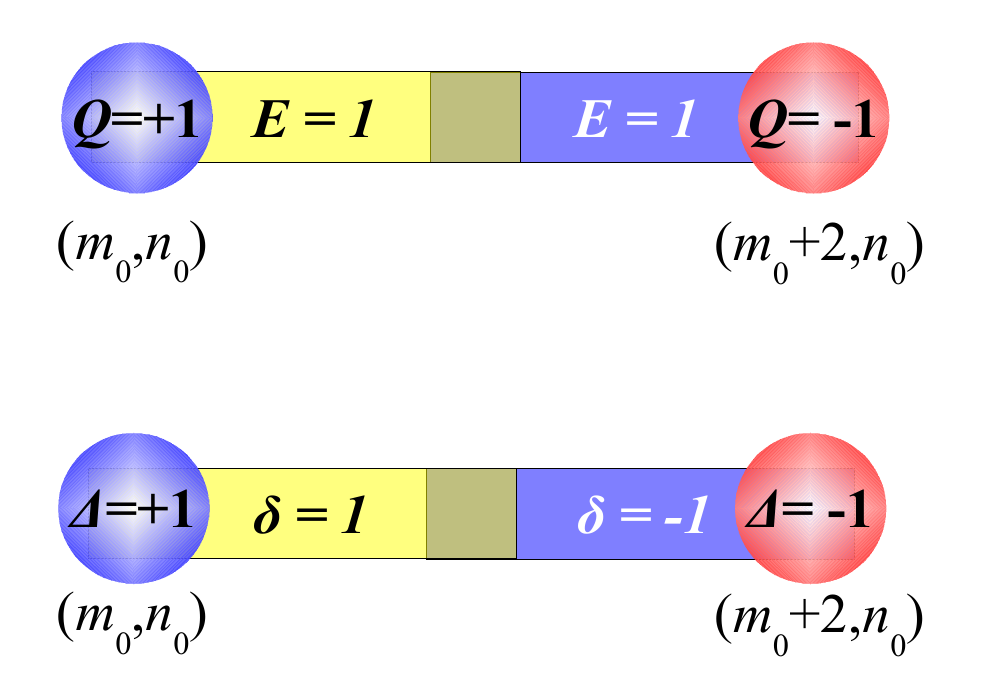}\\
  \caption{An example of the charge and flux configurations, for R=2 . The different colors represent the condensate species.
  The upper couple of charges are with QED quantum numbers, and the lower couple with BEC local number deviations quantum numbers.
  Such a 'flux tube' can be embedded, in absence of other charges, in a lattice whose other links carry $E=0$.}\label{fig2}
\end{figure}

Denote $\left|R\right\rangle $ as the state of two such 'external
charges', $Q_{m_{0},n_{0}}=1,Q_{m_{0}+R,n_{0}}=-1$.
In the strong coupling limit, it can be written as a perturbative
series, whose zeroth order term is
$\left|R_{0}\right\rangle =\underset{m_{0}\leq m<m_{0}+R}{\prod}\widetilde{a}_{m,n}^{\dagger}\left|\left\{ 0\right\} \right\rangle$.
This corresponds to a 'flux tube' from a positive charge to a negative one.
Thus, in the strong limit, we get, indeed, the expected strong coupling
linear behavior of the energy,
\begin{equation}
\overline{E}\left(R\right)\equiv\frac{1}{U_{0}}E\left(R\right)=\frac{g^{2}}{2}R+O\left(g^{-6}\right)
\end{equation}

The effect can then be observed by measuring the local density deviations
$\delta_{m,n}^{k}$, which are expected, in the leading order, to
have a magnitude 1 and alternating signs between the two 'charges'.
An example for $R=2$ can be seen in figure \ref{fig2}.
When $R$ is too large, the energy of $H_{G}$ is smaller than the energy of such a 'flux tube',
and then the low energy theory breaks, and a 'flux tube' is no longer the state of minimal energy.
The low energy picture holds as long as the 'flux tube' length satisfies $R\ll\lambda / \mu$.

Outside the strong coupling regime, such perturbative calculations are no longer valid. However, in 2+1 dimensions
the confinement holds for all values of $g$. Hence, the effect should be seen experimentally even slightly outside the strong coupling limit, although not in the weak limit (in order to fit with the Quantum Rotor approximation).

It may also be possible to experimentally observe the effect of a finite temperature, $T>0$,
on the model, including a phase transition.

\emph{Extensions of the model.} In this paper, we have shown a method to simulate compact QED using BECs in optical lattices,
as a way to observe charge confinement. The suggested model can be extended in several ways: More realistic,
nonperiodic boundary conditions can be imposed (e.g., no 'charges' on the boundary); Using additional condensates (of new species), a 3+1 simulations
could be achieved. Interestingly, a dynamical charge which is minimally coupled to the field can be implemented using another condensate species.
This is equivalent to a special case of the model of Fradkin and Shenker \cite{FradkinShenker}, in which a Higgs field
with a 'frozen' radius is coupled to a U(1) gauge field. \cite{futurework}.

To Conclude, we hope that this model could serve as one of the building blocks of the bridge into the world of dynamic quantum gauge field theories simulations.

\emph{Acknowledgments.} The authors would like to thank A. Casher, J. I. Cirac, S. D\"{u}rr,
O. Kenneth, S. Nussinov and B. Svetitsky for helpful discussions.
This work has been supported by the Israel Science Foundation, the German-Israeli
Foundation, and the European Commission (PICC).

\bibliography{ref}

\end{document}